\documentclass[11pt]{article}
\usepackage{latexsym}
\usepackage[dvips]{graphics,color}
\usepackage{graphpap}
\usepackage{alltt}
\usepackage{graphicx}
\usepackage{amsfonts}
\usepackage{amsmath}
\usepackage{amssymb}
\usepackage{psfrag}
\newtheorem{theorem}{Theorem}[section]
\newtheorem{lemma}[theorem]{Lemma}
\newtheorem{proposition}[theorem]{Proposition}

\newtheorem{definition}[theorem]{Definition}

\newcommand{\beq}{\begin{equation}}
\newcommand{\feq}[1]{\label{#1} \end{equation}}
\newcommand{\beqr}{\begin{eqnarray}}
\newcommand{\feqr}{\end{eqnarray}}
\def\non{\nonumber}

\newcommand{\rf}[1]{(\ref{#1})}

\definecolor{red}{rgb}{1,0,0}
\DeclareFontFamily{U}{eufm}{}
\DeclareFontShape{U}{eufm}{m}{n}{<->eufm10}{}
\DeclareSymbolFont{mcy}{U}{eufm}{m}{n}
\DeclareMathSymbol{\Hr}{\mathord}{mcy}{"58}

\usepackage[active]{srcltx}
\begin{document}

\begin{center}

%\vspace{2cm}

{\Large \bf Self-similar solutions of R\'{e}nyi's entropy and the concavity of its entropy power}\\
[4mm]

\large{Agapitos N. Hatzinikitas}\footnote{On leave of absence from: Department of Mathematics, University of Aegean, School of Sciences, Karlovasi, 83200, Samos, Greece.} \\ [5mm]

{\small School of Physics and Astronomy, \\ 
University of Leeds, \\
Leeds, LS2 9JT,\\
United Kingdom\\
E-mail: ahatz@aegean.gr}\\ [5mm]
\end{center}

\begin{abstract}
We study the class of self-similar probability density functions with finite mean and variance which maximize R\'{e}nyi's entropy. The investigation is restricted in the Schwartz space $S(\mathbb{R}^d)$ and in the space of $l$-differentiable compactly supported functions $C_c^l(\mathbb{R}^d)$. Interestingly the solutions of this optimization problem do not coincide with the solutions of the usual porous medium equation with a Dirac point source, as it occurs in the optimization of Shannon's entropy. We also study the concavity of the entropy power in $\mathbb{R}^d$ with respect to time using two different methods. The first one takes advantage of the solutions determined earlier while the second one is based on a setting that could be used for Riemannian manifolds.
\end{abstract}

%%%%%%%%%%%%%%%%%%%%%%%%%%%%%%%%%%%%%%%%%%%%%%%%%%%%%%%%%%%%%%%%%%%%%%%%%%%%%%%%%%%%%%%%%%%%%%%%%%
%%%%%%%%%%%%%%%%%%%%%%%%%%%%%%%%%%%%%%%%%%%%%%%%%%%%%%%%%%%%%%%%%%%%%%%%%%%%%%%%%%%%%%%%%%%%%%%%%%
\noindent\textit{MSC2010:} 94A17, 60E99 \\
\textit{Key words:} Maximum R\'{e}nyi entropy, Entropy power, Fisher information, Nonlinear diffusion equation  
%%%%%%%%%%%%%%%%%%%%%%%%%%%%%%%%%%%%%%%%%%%%%%%%%%%%%%%%%%%%%%%%%%%%%%%%%%%%%%%%%%%%%%%%%%%%%%%%%%
%%%%%%%%%%%%%%%%%%%%%%%%%%%%%%%%%%%%%%%%%%%%%%%%%%%%%%%%%%%%%%%%%%%%%%%%%%%%%%%%%%%%%%%%%%%%%%%%%%
\section{Introduction}
\label{sec0}
The last two decades have witnessed an enormous  growing interest in using information concepts in diverse fields of science. Although R\'{e}nyi entropy was introduced as early as 1961, only recently a wide range of applications has emerged as in the analysis of quantum entanglement \cite{HO}, quantum correlations \cite{ES}, computer vision \cite{DG}, clustering \cite{RJ}, quantum cryptography \cite{CB} and pattern recognition \cite{PS}.
\par In the present work we solve three problems. The first examines the possibility to extremize R\'{e}nyi's entropy using self-similar probability density functions (p.d.f.'s) with zero expectation value and finite second moment in the Schwartz space and the space of compactly supported continuous functions on $\mathbb{R}^d$. The second tackles the same problem but with the additional feature of non-zero mean. Finally, the third one is devoted to the determination of conditions under which the concavity of entropy power is valid.
\par Our contribution is threefold: First, we theoretically establish the solutions of the first two problems by applying the method of calculus of variations, which was lacking from the literature. Second, we compare the specified solutions with the already known ones derived from the fast and porous medium equations. Third, we propose two different methods to answer the third problem.
\par In particular, for the first problem, the functional which contains R\'{e}nyi's entropy and the constraints, incorporated as Lagrange multipliers, is constructed and then by applying the calculus of variations its critical points is determined. The perturbed p.d.f.'s have the form $g_{\epsilon}=f+\epsilon h$ where $|\epsilon|<\epsilon_0<1$ and $h$ are functions which are chosen in such a way that $g_{\epsilon}$ is a p.d.f. and has the same variance as $f$. The vanishing of the first variation of the Lagrange functional provides the equation which determines the critical points. It  turns out that the solutions are unique. The nonnegativity of the second variation leads to an integral inequality which is preserved by the admissible perturbations we consider. Therefore the critical point is a local maximum of the functional. To prove its global nature we use the concept of the relative R\'{e}nyi entropy and examine its positivity at the critical point. This procedure can be generalised in $\mathbb{R}^d$ and also by requiring a finite covariance constraint the well-known solutions of \cite{JV} are recovered. As a check one can prove that in the $\alpha\rightarrow 1$ limit the solutions converge to the p.f.d. of the normal distribution $\mathcal{N}(0,\mu_2)$. The second problem is proved to be equivalent to the first one by performing a suitable transformation to the random variable (composition of a displacement with a rescaling). Therefore its solution maps to the one of the first problem. 
\par The knowledge of the solutions enables us to construct the nonlinear diffusion equation they satisfy and compare them with those of Zel'dovich, Kompaneets and Barenblatt (ZKB) \cite{B}, \cite{ZK}. The difference is in the diffusion coefficient which depends not only on the shape and the size of a molecule but also on the order $\alpha$ of the R\'{e}nyi entropy and the dimension of the space. We plot both our solution and the Barenblatt's solutions and observe that their $\mathcal{L}_{\infty}$ norms obey $\|u\|_{\infty}\leq\|f\|_{\infty}, \, \forall \frac{d}{d+2}<\alpha \leq \alpha_{\textrm{th.}}(d)$ while for values of $\alpha$ greater than the threshold $\alpha_{\textrm{th.}}(d)$ the inequality changes direction.
\par Finally, the problem of concavity of the entropy power in $\mathbb{R}^d$ is confronted by utilising two different methods. In the first method, the solutions of the first problem guarantee concavity on the condition that the second time derivative of R\'{e}nyi's entropy fulfils inequality \rf{sec5 : eq3}. The second method is closer to the spirit analyzed in \cite{V}, where the $\alpha=1$ case was studied, and concavity holds provided that \rf{sec5 : eq17} is satisfied.       
\par The paper is organized into six sections. Section 2 reviews and proves some properties of the entropy. Section 3 determines the solutions of the two maximization problems using the method of calculus of variations and examines their global validity using the concept of relative R\'{e}nyi entropy. Section 4 provides the nonlinear diffusion equation the solutions satisfy and compare it with the usual one. Section 5 proves the concavity of the R\'{e}nyi entropy with respect to time following two different methods. Section 6 concludes the work and comments on more general constraints one could have considered.  

%%%%%%%%%%%%%%%%%%%%%%%%%%%%%%%%%%%%%%%%%%%%%%%%%%%%%%%%%%%%%%%%%%%%%%%%%%%%%%%%%%%%%%%%%%%%%%%%%%%%%%%%%%%%
\section{Preliminaries}
\label{sec1}
In this section we briefly review some properties of the R\'{e}nyi entropy and for the sake of completeness we present the corresponding proofs.
\begin{definition}
Let $(\Omega, \mathcal{A},\mu)$ be a probabilty space and an $\mathcal{A}$-measurable function $f: \Omega \rightarrow \mathbb{R}^+$ be a probability density function (p.d.f.). The differential R\'{e}nyi entropy of order $\alpha$, $\alpha \in \mathbb{R}^+$,  is the nonlinear functional 
\beqr
H_{\alpha}: f \rightarrow \mathbb{R}
\label{sec1 : eq1}  
\feqr  
defined by
\beqr
H_{\alpha}[f]:=\frac{1}{1-\alpha}\ln\left(\int_{\Omega}\left(\frac{dF}{d\mu}\right)^{\alpha-1}dF\right)
\label{sec1 : eq2}
\feqr
where $F$ is the probability measure induced by $f$ namely
\beqr
F(E)=\int_E f(x)d\mu(x), \quad \forall E\in \mathcal{A}.
\label{sec1 : eq3}
\feqr
\end{definition}
Other equivalent ways of defining the R\'{e}nyi entropy are
\beqr
H_{\alpha}[f]&=& \frac{1}{1-\alpha}\ln\left(\mathbb{E}_X (f^{\alpha -1}(X)) \right), \quad \forall \alpha \non \\
H_{\alpha}[f]&=& \frac{1}{1-\alpha}\ln\left(\left\|f\right\|_{L^{\alpha}}^{\alpha} \right), \quad \forall \alpha>1. \non
\feqr
\textbf{Properties}
\begin{enumerate}
\item[($1\alpha$)] $H_{\alpha}[f]$ is continuous ($\alpha\neq1$) and strictly decreasing function in $\alpha$ unless f is the uniform density in which case it is constant. 
\item[($1\beta$)] A consequence of property ($1\alpha$) is the inequality
\beqr
&&0<D_{KL}(g||f)<H_1[g]-(2-\alpha)H_{\alpha}[f], \quad \textrm{for} \quad \alpha<1 \label{sec1 : eq4} \\
&& \textrm{where} \quad g=\frac{f^{\alpha}}{\int_{\Omega}f^{\alpha}d\mu}, \quad \int_{\Omega}(g-f)d\mu>0 \quad \textrm{and} \quad D_{KL}(g||f) =\int_{\Omega}g\ln\left(\frac{g}{f}\right)d\mu \non
\feqr
is the Kullback-Leibler (KB) relative entropy. If $\alpha >1$ then the direction of the inequality is reversed. Therefore the R\'{e}nyi entropy as a function of $\alpha$, for fixed $f$, is bounded by the difference between the Shannon entropy of $g$ and the KB relative entropy of $f$ and $g$.\\
\textbf{Proof}
\begin{enumerate}
\item[($1\alpha$)] By H\"{o}lder's inequality there is a family of relations
\beqr
\int_{\Omega}f^{(1-\theta)p+\theta q} d\mu \leq \left(\int_{\Omega}f^{p} d\mu\right)^{1-\theta}\left(\int_{\Omega}f^{q} d\mu\right)^{\theta}
\label{sec1 : eq5}
\feqr
holding whenever $p,q\geq 0$ and $\theta\in [0,1]$. Taking $p=1$ and assuming $f$ to be a p.d.f. we have
\beqr
\int_{\Omega}f^{1-\theta+\theta q} d\mu \leq \left(\int_{\Omega}f^{q} d\mu\right)^{\theta}
\label{sec1 : eq6}
\feqr
Let $0<\alpha<\beta<1$ and $q=\alpha$. Then for $\theta=(1-\beta)/(1-\alpha)<1$ the previous inequality becomes
\beqr
\int_{\Omega}f^{\beta} d\mu \leq \left(\int_{\Omega}f^{\alpha} d\mu\right)^{\frac{1-\beta}{1-\alpha}} 
\label{sec1 : eq7}
\feqr 
which, using that the $\ln$-function is an increasing function, implies that
\beqr
H_{\beta}[f]\leq H_{\alpha}[f]. \non
\feqr
\end{enumerate} 
The same proof holds for $\alpha, \beta >1$. \\
\item[($1\beta$)] Differentiating $H_{\alpha}$ with respect to $\alpha$ we obtain
\beqr
\frac{dH_{\alpha}[f] }{d\alpha}&=&\frac{1}{(1-\alpha)}\left(H_{\alpha}[f]- \frac{1}{\alpha \int_{\Omega} f^{\alpha}(x) d\mu} H_1[f^{\alpha}]\right] \non \\
&=& \frac{1}{\alpha(1-\alpha)}\left(H_{\alpha}[f]+ D_{KL}(g||f)-\frac{1}{\alpha}H_1[f^{\alpha}]\right) \non \\
&=& \frac{1}{\alpha(1-\alpha)}\left((2-\alpha)H_{\alpha}[f]+D_{KL}(g||f)-H_1[g]\right)<0, \,\, \forall \alpha<1 
\label{sec1 : eq8}
\feqr
from which the inequality follows.
\item[($2$)] $H_{\alpha}(f)$ as a function of $\alpha$ converges to the following limits
\beqr
\lim_{\alpha \rightarrow 0^+}H_{\alpha}[f]&=&\ln(\mu(\Omega)) \label{sec1 : eq9} \\
\lim_{\alpha \rightarrow 1}H_{\alpha}(f)&=&H_1[f]=-\int_{\Omega} f \ln f d\mu \label{sec1 : eq10} \\
\lim_{\alpha \rightarrow \infty}H_{\alpha}[f]&=&\left\| f \right\|_{\infty}=\textrm{inf.} \{0<\lambda<\infty: \,\, f\leq \lambda \,\, \textrm{a.e.} \,\, \omega \in \Omega\}.
\label{sec1 : eq11}
\feqr
where $H_1[f]$ is the Shannon entropy of f.
\item[($3$)] Let $f$ be a non-negative and integrable function w.r.t. the measure $\mu$ on $\Omega$, then the following inequality holds
\beqr
H_{\alpha}[f] < \left(\frac{\alpha}{1-\alpha}\right) \ln \left(\int_{\Omega} f d\mu\right) +\ln(\mu(\Omega)) \quad \textrm{for} \quad \alpha<1 
\label{sec1 : eq12}
\feqr
while for $\alpha>1$ the inequality is reversed.\\
\textbf{Proof}\\
This is a direct consequence of Jensen's inequality 
\beqr
\frac{1}{\mu(\Omega)}\int_{\Omega} (\phi \circ f) d\mu \leq \phi \left(\frac{1}{\mu(\Omega)}\int_{\Omega} f d\mu \right)
\label{sec1 : eq13} 
\feqr
for concave functions $\phi$. In our case $(\phi\circ f)(x))=f(x)^{\alpha}$ is concave for $\alpha<1$ and by applying it we have
\beqr
\frac{1}{1-\alpha} \ln \left(\frac{1}{\mu(\Omega)}\int_{\Omega} f^{\alpha} d\mu\right) < \left(\frac{\alpha}{1-\alpha}\right) \ln \left( \frac{1}{\mu(\Omega)}\int_{\Omega} f d\mu\right)
\label{sec1 : eq14}
\feqr
from which it is deduced straightforwardly. In particular for a p.d.f. it reduces to $H_{\alpha}[f] <\ln(\mu(\Omega))$.
\item[($4$)] If the $L_1(\Omega, \mathbb{R}^+)$ norm is invariant under the homogeneous dilations
\beqr
x&\rightarrow& \tilde{x}=\lambda^{\gamma} x, \quad \lambda>0, \,\, \forall x\in \Omega \subset \mathbb{R}^d \non \\  
f(x)&\rightarrow& f_{\lambda}(x)= \lambda^{\delta} f(\tilde{x}) 
\label{sec1 : eq15}
\feqr
then $H_{\alpha}$, for $\alpha<1$, scales as
\beqr
H_{\alpha}[f_{\lambda}]=H_{\alpha}[f]-\delta \ln \lambda, \quad \delta=d\gamma.
\label{sec1 : eq16}
\feqr
\textbf{Proof}
The $L_1$ norm invariance of $f$ 
\beqr
\left\| f \right\|_{L_1(\Omega)}=  \left\| f \right\|_{L_1(\tilde{\Omega})}=M
\label{sec1 : eq17}
\feqr
implies the condition 
\beqr
\delta=d\gamma
\label{sec1 : eq18}
\feqr
which combined with the definition of R\'{e}nyi's entropy produces the desired result.
\end{enumerate}

%%%%%%%%%%%%%%%%%%%%%%%%%%%%%%%%%%%%%%%%%%%%%%%%%%%%%%%%%%%%%%%%%%%%%%%%%%%%%%%%%%%%%%%%%%%%%%%
%%%%%%%%%%%%%%%%%%%%%%%%%%%%%%%%%%%%%%%%%%%%%%%%%%%%%%%%%%%%%%%%%%%%%%%%%%%%%%%%%%%%%%%%%%%%%%%%
\section{Formulation of the first problem and its solutions}
\label{sec2}
In what follows we restrict on the probability space $(\mathbb{R}, \mathcal{B}, dx)$ where $\mathcal{B}=\sigma(\mathcal{O})$ is the sigma algebra on open sets and $dx$ the Lebesgue measure  on $\mathbb{R}$. The domain of the R\'{e}nyi functional, $\mathcal{D}(H_{\alpha})$, is defined to be
\beqr
\mathcal{D}(H_{\alpha<1})&=&\left\{ f\in S(\mathbb{R}): \int_{\mathbb{R}} f(x)dx=1\right\} \label{sec2 : eq1} \\
\mathcal{D}(H_{\alpha>1})&=&\left\{ f_+\in C_c^k(\mathbb{R}), \, k\in \mathbb{Z}^+ \,\,\textrm{and} \,\,  k\leq \left[\frac{1}{\alpha-1}\right]: \int_{\mathbb{R}} f_+(x)dx=1\right\},
\label{sec2 : eq2}
\feqr
where $f$ is a positive and integrable real valued function, $f_+(x)=f(x) \chi_{B}(x)$ with $\chi$ the indicator function of the set $B$ and $[\cdot]$ is the integer part of the number. 
\par The first entropy maximization problem with vanishing mean and finite variance is formulated as:
\beqr
\underset{f\in \mathcal{D}(H_{\alpha<1})}{max. H_{\alpha}[f]} \quad \textrm{subjected to the constraint} \quad \mathbb{E}(X^2)=\int_{\mathbb{R}} x^2 f(x) dx =\mu_2^2<\infty.
\label{sec2 : eq3}
\feqr
Using the method of Lagrange multipliers we construct the functional
\beqr
\mathcal{F}(f; \lambda_0,\lambda_2)=H_{\alpha}[f]-\sum_{k\in \{0,2\}} \lambda_k \left(\mathbb{E}(X^k) -\mu_k^2\right), \quad \mu_0=1
\label{sec2 : eq4}
\feqr
and impose appropriate conditions on the  perturbations in order to calculate its first and second variation.  
\begin{definition}
A perturbation $h\in S(\mathbb{R})$ is called admissible if it satisfies the following conditions:
\beqr
|h(x)|&<&c |f(x)|,\,\, 0<c<1, \,\, \forall x\in \mathbb{R}, \non \\
\int_{\mathbb{R}} h(x) dx &=& 0 \quad \textrm{and} \quad \int_{\mathbb{R}} x^2 h(x) dx =0. 
\label{sec2 : eq5}
\feqr 
\end{definition}
If we introduce the usual inner product in $S(\mathbb{R})$, the previous integral conditions imply that we search for a class of functions which are orthogonal to the unity and $x^2$. Odd functions in the Schwartz space such as $Q(x)e^{-b|x|^{\mu}}, \, \, b, \, \mu>0$ with $Q(x)$ be a polynomial of odd powers of x satisfies these criteria.
\par Expanding the Lagrange functional in a Taylor series up to second order in $\epsilon$ we obtain
\beqr
\mathcal{F}(f+\epsilon h; \lambda_0,\lambda_2)=\mathcal{F}(f; \lambda_0,\lambda_2)+\delta \mathcal{F}_{f}(h) \epsilon+\delta^2 \mathcal{F}_{f}(h) \epsilon^2+o(\epsilon^2).
\label{sec2 : eq6}
\feqr
The first-order necessary condition for optimality requires  \cite{DL}
\beqr
\delta \mathcal{F}_{\hat{f}}(h)&=&\lim_{\epsilon \rightarrow 0^+} \frac{\mathcal{F}(f+\epsilon h; \lambda_0,\lambda_2)-\mathcal{F}(f; \lambda_0,\lambda_2)}{\epsilon} \non \\
&=&\frac{d}{d\epsilon}\mathcal{F}(f+\epsilon h; \lambda_0,\lambda_2)\biggr|_{\epsilon=0}=0, \quad |\epsilon|<\epsilon_0<1
\label{sec2 : eq7}
\feqr 
or, equivalently,
\beqr
\int_{\mathbb{R}} \biggl(f^{\alpha -1}(x)-\sum_{k\in \{0,2\}} \tilde{\lambda}_k x^k \biggr) h(x) dx=0, \quad \textrm{where} \quad \tilde{\lambda}_k=\lambda_k\, g(\alpha)=\lambda_k\frac{(1-\alpha)}{\alpha}\int_{\mathbb{R}} f^{\alpha}(x) dx .
\label{sec2 : eq8}
\feqr
The function $f(x)$ is determined by using the following lemma 
\begin{lemma}
If $f \in S(\mathbb{R})$ and if
\beqr
\int_{\mathbb{R}} f(x) g(x) dx=0, \,\, \forall g\in S(\mathbb{R}) \quad \textrm{such that} \quad \lim_{|x|\rightarrow \infty}g(x)=0 
\label{sec2 : eq9}
\feqr
then $f(x)=0, \,\, \forall x\in \mathbb{R}$.
\end{lemma}
\textbf{Proof} \\
Suppose that $f(x)\neq 0$ then there exists $\xi \in \mathbb{R}$ such that $f(\xi)=c>0$ (assuming that the constant is positive). Since $f \in S(\mathbb{R})$ there exists a neighbour $(a,b)$ of $\xi$ in which $f(x)>0, \, \forall x\in (a,b)$. Define the function
\beqr
g(x)=(x-a)(b-x)e^{-x^2} \chi_{(a,b)}(x).
\label{sec2 : eq10}
\feqr
The function $g(x)$ is positive in $(a,b)$ with $g(a)=g(b)=0$. However
\beqr
\int_{\mathbb{R}}f(x)g(x)dx=\int_a^b f(x)(x-a)(b-x)e^{-x^2}dx>0
\label{sec2 : eq11}
\feqr
since the integrand is positive (except at a and b). This contradiction proves the lemma. 
 \\
Therefore $f(x)$ is given by 
\beqr
\hat{f}(x)=\biggl(\sum_{k\in \{0,2\}} \tilde{\lambda}_k x^k \biggr)^{-\frac{1}{1-\alpha}}=\tilde{\lambda}_0^{-\frac{1}{1-\alpha}}  \left(1+\frac{\tilde{\lambda}_0}{\tilde{\lambda}_2} x^2 \right)^{-\frac{1}{1-\alpha}}=\left\| \hat{f} \right\|_{\infty} \left(1+\frac{\tilde{\lambda}_0}{\tilde{\lambda}_2} x^2 \right)^{-\frac{1}{1-\alpha}},
\label{sec2 : eq12}
\feqr
where the $L^{\infty}$ norm is with respect to x and $\tilde{\lambda}_k>0$ since $\hat{f}\in \mathcal{D}_{\alpha<1}$. \\
In order $\hat{f}$ to be a local maximum the following second-order necessary condition for optimality should also hold
\beqr
\delta^2 \mathcal{F}|_{\hat{f}}(h)\leq 0 \quad \textrm{or} \quad \frac{d^2}{d\epsilon^2}\mathcal{F}(f+\epsilon h; \lambda_0,\lambda_2)\biggr|_{\epsilon=0, f=\hat{f}}\leq 0.
\label{sec2 : eq13}
\feqr
In other words, the second variation of $\mathcal{F}$ at $\hat{f}$ should be positive semidefinite on the space of admissible perturbations $h$. The previous inequality is translated into 
\beqr
\frac{\alpha}{\alpha-1}\int_{\mathbb{R}} \hat{f}^{\alpha -1}(x) h(x) dx \leq \left(\int_{\mathbb{R}} \hat{f}^{\alpha}(x) dx\right) \left(\int_{\mathbb{R}} \hat{f}^{\alpha -2}(x) h^2(x) dx\right).
\label{sec2 : eq14}
\feqr
It is easily checked that the solution $\hat{f}$ together with the admissible perturbations satisfy the strict inequality, therefore $\hat{f}$ is a strict one-parameter family of local maxima with increasing R\'{e}nyi entropy as property $(1\alpha)$ of section 2 guarantees.\\  
\textbf{Remark}
\begin{description}
\item If $0<\int_{\mathbb{R}}f^{\alpha}(x)dx\leq 2$ then Taylor expanding the $\ln$-function around unity, the R\'{e}nyi entropy reduces to the Havrda-Charv\'{a}t entropy \cite{HC} also called Tsallis entropy \cite{CT} 
\beqr
S_{\alpha}(f)=\frac{1}{\alpha-1}\left(1-\int_{\mathbb{R}}f^{\alpha}(x)dx \right).
\label{sec2 : eq15}
\feqr
The solution in this case is the previous one but with the substitution $\tilde{\lambda}_k=\lambda_k\frac{(1-\alpha)}{\alpha}$. 
\end{description}
Depending on the space of functions, we distinguish the following two types of solutions.
\begin{enumerate}
\item[\textbf{$(3\alpha)$}]{\textbf{The $\boldsymbol{S(\mathbb{R}})$ solution ($\boldsymbol{\alpha<1}$).}}
The requirement that $\hat{f}$ is a p.d.f. leads to 
\beqr
\tilde{\lambda}_0^{\frac{1}{2}-\frac{1}{1-\alpha}} B_{\frac{1}{2}}=\tilde{\lambda}_2^{\frac{1}{2}} \quad \textrm{for} \quad 0<\alpha<1,
\label{sec2 : eq16}
\feqr
while the second moment constraint gives
\beqr
\tilde{\lambda}_0^{\frac{3}{2}-\frac{1}{1-\alpha}} B_{\frac{3}{2}}=\tilde{\lambda}_2^{\frac{3}{2}}\mu_2^2(\alpha) \quad \textrm{for} \quad \frac{1}{3}<\alpha<1.
\label{sec2 : eq17}
\feqr
We adopt the abbreviation $B_{s/2}\equiv B(s/2,1/(1-\alpha)-s/2)$ from now on. From these two relations we conclude that
\beqr
\frac{\tilde{\lambda}_0}{\tilde{\lambda}_2} \frac{B_{\frac{3}{2}}}{B_{\frac{1}{2}}}=\frac{\lambda_0}{\lambda_2} \left(\frac{1-\alpha}{3\alpha -1}\right)=\mu_2^2 \quad \textrm{for} \quad \frac{1}{3}<\alpha<1.
\label{sec2 : eq18}
\feqr
Using this relation, the exact solution can be expressed as
\beqr
\hat{f}(x)=\left(\frac{1-\alpha}{3\alpha-1}\right)^{\frac{1}{2}}\frac{1}{\mu_2 B(\frac{1}{2},\frac{1}{1-\alpha}-\frac{1}{2})}\left(1+\left(\frac{1-\alpha}{3\alpha-1}\right) \frac{x^2}{\mu_2^2}\right)^{-\frac{1}{1-\alpha}}, \quad \frac{1}{3}<\alpha<1.
\label{sec2 : eq19}
\feqr
This one-parameter family of local maxima of $H_{\alpha}$ is unique and it remains to prove that it is actually also a one-parameter family of global maxima in $\mathcal{D}(H_{\alpha<1})$. For this we use the notion of the relative $\alpha$-R\'{e}nyi entropy of two densities $\hat{f}$ and $g$, defined by \cite{LYZ} 
\beqr
D_{\alpha}(g||\hat{f})=\frac{1}{1-\alpha}\ln \left(\int_{\mathbb{R}} \hat{f}^{\alpha-1}(x) g(x) \, dx \right) +\frac{(1-\alpha)}{\alpha} H_{\alpha}[\hat{f}] -\frac{1}{\alpha} H_{\alpha}[g],
\label{sec2 : eq20}
\feqr
where $g$ satisfies the same second moment constraint as $\hat{f}$. The first term in the right hand side of \rf{sec2 : eq20} equals to $H_{\alpha}[\hat{f}]$ as one may check since $\int_{\mathbb{R}} \hat{f}^{\alpha-1}(x) g(x) dx=\int_{\mathbb{R}} \hat{f}^{\alpha}(x)dx $. Therefore
\beqr
D_{\alpha}(g||\hat{f})=\frac{1}{\alpha}(H_{\alpha}[\hat{f}]-H_{\alpha}[g]) \geq 0
\label{sec2 : eq21}
\feqr
by applying H\"{o}lder's inequality to the functions $(\hat{f}^{\alpha-1}g)^{\alpha}$ and $\hat{f}^{\alpha(\alpha-1)}$. The same result holds in the $\alpha>1$ case.
\begin{proposition}
The optimization problem for $\alpha<1$ has a one-parameter family of global maxima $\hat{f}$ which in the $\alpha \rightarrow 1^{-}$ limit converges to the global maximum of the Shannon entropy which is the normal distribution $\mathcal{N}(0,\mu_2)$.\\
\end{proposition}
\textbf{Proof}\\
The first two terms of \rf{sec2 : eq19} in the $\alpha \rightarrow 1^-$ limit give
\beqr
\lim_{\alpha \rightarrow 1^{-}} \tilde{\lambda}_0^{-\frac{1}{1-\alpha}}(\alpha)=\frac{1}{\sqrt{2\pi \mu_2^2}},
\label{sec2 : eq22}
\feqr
while the third term, by performing the change of variables $s=(1-\alpha)/(3\alpha-1), \rho=1/2s$ successively, converges to
\beqr
\lim_{\alpha \rightarrow 1^{-}}\left(1+\left(\frac{1-\alpha}{3\alpha-1}\right) \frac{x^2}{\mu_2^2}\right)^{-\frac{1}{1-\alpha}}&=&\lim_{s\rightarrow 0^{+}} \left(1+s\frac{x^2}{\mu_2^2}\right)^{-\frac{3}{2}-\frac{1}{2s}} \non \\
&=&\lim_{\rho \rightarrow 0^{+}}\left(1+\frac{x^2}{2\mu_2^2}\rho\right)^{-\frac{1}{\rho}} \non \\
&=& e^{-\frac{x^2}{2\mu_2^2}}.
\label{sec2 : eq23}
\feqr 
\item[$(3\beta)$]{\textbf{The l-differentiable compactly supported solution ($\boldsymbol{\alpha>1}$).}}  
In this case following steps similar to the previous one the solution turns out to be
\beqr
\hat{f}(x)=\tilde{\lambda}_0^{\frac{1}{\alpha-1}}\left(1-\left(\frac{\alpha-1}{3\alpha-1}\right)\frac{x^2}{\mu_2^2}\right)_{+}^{\frac{1}{\alpha-1}}, \,\,\, \alpha>1\quad \textrm{and} \quad x_{+}\equiv max\{0,x\}\equiv x\theta(x),
\label{sec2 : eq24}
\feqr
where $\theta(x)$ is the unit-step function and the arbitrary constant $\tilde{\lambda}_0$ is specified as usual by imposing the requirement that $\hat{f}$ be a p.d.f.
\beqr
\tilde{\lambda_0}(\alpha)=\left(\left(\frac{\alpha-1}{3\alpha-1}\right)^{\frac{1}{2}} \frac{1}{\mu_2 B(\frac{1}{2},\frac{1}{\alpha-1}+1)}\right)^{\alpha-1}.
\label{sec2 : eq25}
\feqr 
\end{enumerate}
\textbf{Remarks}\\
\begin{description}
\item{$\bullet$} The solution $(3\alpha)$ can also be derived by integrating equation \rf{sec2 : eq8}, in which case we obtain the relation between the Lagrange multipliers
\beqr
\lambda_0=\frac{\alpha}{1-\alpha}-\mu_2^2 \lambda_2.
\label{sec2 : eq26}
\feqr
Eliminating $\lambda_0$ from the solution and applying the constraints we arrive at \rf{sec2 : eq19}.
\item{$\bullet$} The derivation of $(3\alpha)$ and $(3\beta)$ is based on \rf{apA : eq1} and \rf{apA : eq2} which are connected through the transformation $u=x^2/(x^2+1)$ with $u\in (0,1)$, see \rf{apA : eq3}. 
\item{$\bullet$} The previous set up can also be applied to the more general case in which a covariance matrix $C$-constraint is present. The new solution can be derived from the old one by replacing $x^2$ by $x^T C^{-1} x $, see \cite{JV}. 
\end{description}
%%%%%%%%%%%%%%%%%%%%%%%%%%%%%%%%%%%%%%%%%%%%%%%%%%%%%%%%%%%%%%%%%%%%%%%%%%%%%%%%%%%%%%%%%%%%%%%
%%%%%%%%%%%%%%%%%%%%%%%%%%%%%%%%%%%%%%%%%%%%%%%%%%%%%%%%%%%%%%%%%%%%%%%%%%%%%%%%%%%%%%%%%%%%%%%%
\section{Formulation of the second problem and its solutions}
\label{sec3}

The second entropy maximization problem with non-vanishing mean and finite variance is formulated as follows:
\beqr
\begin{array}{lc}\underset{f_X\in \mathcal{D}(H_{\alpha<1})}{max. H_{\alpha}[f_X]} \quad \textrm{subjected to the constraints} & \mathbb{E}(X)=\int_{\mathbb{R}} x f_X(x) dx =\mu_1<\infty \\
& \textrm{Var}(X)=\mathbb{E}(X^2)-\mu_1^2=\mu_2^2<\infty.\end{array}
\label{sec3 : eq1}
\feqr
This problem can equivalently be restated as
\beqr
\begin{array}{lc}\underset{f_Y\in \mathcal{D}(H_{\alpha<1})}{max. H_{\alpha}[f_Y]} \quad \textrm{subjected to the new constraints} & \mathbb{E}(Y)=\int_{\mathbb{R}} y f_Y(y) dy =0 \\
& \textrm{Var}(Y)=\mathbb{E}(Y^2)=1.\end{array}
\label{sec3 : eq2}
\feqr
The random variables $X, Y$ are related through the transformation $X=\mu_2 Y+\mu_1$ and their corresponding p.d.f.'s $f_X, f_Y$ satisfy
\beqr
f_X(x)=\frac{1}{\mu_2}f_Y\left(\frac{x-\mu_1}{\mu_2}\right).
\label{sec3 : eq3}
\feqr
As a consequence the entropies are given by
\beqr
H_{\alpha}[f_X]=H_{\alpha}[f_Y]+\ln \mu_2.
\label{sec3 : eq4}
\feqr 
The solutions of the second problem \rf{sec3 : eq1} are therefore given by the solutions of first problem \rf{sec2 : eq3} with the substitution $(x-\mu_1)/\rightarrow x$.   
\par One may also try to solve directly the second problem starting from the Lagrange functional
\beqr
\mathcal{F}(f; \lambda_0, \lambda_1,\lambda_2)=H_{\alpha}[f]-\lambda_0 \left(\int_{\mathbb{R}}f(x)dx -1\right)-2\lambda_1(\mathbb{E}(X)-\mu_1)-\lambda_2(\textrm{Var}(X)-\mu_2^2), \,\, \lambda_k\in\mathbb{R}.
\label{sec3 : eq5}
\feqr
The first-order necessary optimization condition dictates the solution 
\beqr
\hat{f}^{\alpha-1}(x)=\sum_{k=0}^2 \tilde{\lambda}_kx^k,
\label{sec3 : eq6}
\feqr
which can also be proved to be a global maximum. 
\par We distinguish the following two classes of solutions.
\begin{enumerate}
\item[\textbf{$(4\alpha)$}]{\textbf{The $\boldsymbol{S(\mathbb{R}})$ solution ($\boldsymbol{\alpha<1}$).}} The positivity of the solution $\hat{f}^{\alpha-1}(x), \, \forall x\in \mathbb{R}$ requires $\tilde{\lambda}_2>0$ and $\tilde{\lambda}_2\tilde{\lambda}_0-(\tilde{\lambda}_1)^2>0$. The p.d.f. and the mean value constraints lead to the condition
\beqr
\mu_1=-\frac{\lambda_1}{\lambda_2},
\label{sec3 : eq7}
\feqr 
while the variance constraint implies
\beqr
\frac{\lambda_0}{\lambda_2}=\frac{(2n-3)!!}{(2n-5)!!}\mu_2^2+\mu_1^2, \quad n=\frac{1}{1-\alpha}\in \mathbb{Z}^+.
\label{sec3 : eq8}
\feqr 
Finally, the solution is written as
\beqr
\hat{f}(x)=\frac{(2n-2)!!}{(2n-3)!!}\sqrt{\frac{(2n-5)!!}{(2n-3)!!}}\frac{1}{\mu_2}\left(1+\frac{(2n-3)!!}{(2n-5)!!} \left(\frac{x-\mu_1}{\mu_2}\right)^2\right)^{-n},
\label{sec3 : eq9}
\feqr
which is identical to the solution derived from the equivalent problem. 
\item[$(4\beta)$]{\textbf{The l-differentiable compactly supported solution ($\boldsymbol{\alpha>1}$).}} In this case the polynomial $\sum_{k}\tilde{\lambda} kx^k$ should be positive between its real roots. This occurs provided that $\tilde{\lambda}_2<0$ and $(\tilde{\lambda}_1)^2-\tilde{\lambda}_2\tilde{\lambda}_0>0$. Using the indicator function $\chi_{(x_1,x_2)}$, with $x_1, x_2$ the roots of the polynomial, we find the previous solution with a relative minus sign between the terms inside the parentheses while the power is now positive.  
\end{enumerate}
%%%%%%%%%%%%%%%%%%%%%%%%%%%%%%%%%%%%%%%%%%%%%%%%%%%%%%%%%%%%%%%%%%%%%%%%%%%%%%%%%%%%%%%%%%%%%%%
%%%%%%%%%%%%%%%%%%%%%%%%%%%%%%%%%%%%%%%%%%%%%%%%%%%%%%%%%%%%%%%%%%%%%%%%%%%%%%%%%%%%%%%%%%%%%%%%
\section{Comparison with the FME and PME solutions}
\label{sec4}

The p.d.f., $\hat{f}$, which maximizes the Shannon entropy under finiteness of the second moment turns out to be identical to the fundamental solution of the diffusion equation with a Dirac point source. It is worth noting that $\hat{f}$ is actually a global maximum of $H_1$. This observation is accidental as one may justify from the study of the corresponding optimization problem for the R\'{e}nyi entropy. In particular, the nonlinear, initial valued problem
\beqr
\partial_t u(x,t)&=&\Delta u^{\alpha}(x,t), \quad (x,t)\in \mathbb{R}^d\times (0,\infty) \non \\
\lim_{t\rightarrow 0^+} \left(\int_{\mathbb{R}^d}u(x,t)dx\right)&=&\delta_0(x)
\label{sec4 : eq1}
\feqr
has the following self-similar solutions \cite{JV}
\beqr
\textrm{FME solution:} \quad u(x,t)_{\alpha<1}&=& t^{-\delta}\left(C_{\alpha<1}+\frac{(1-\alpha)\delta}{2\alpha d} \left(\frac{x}{t^{\gamma}}\right)^2\right)^{-\frac{1}{1-\alpha}},  \label{sec4 : eq2} \\
\textrm{PME solution:} \quad u(x,t)_{\alpha>1}&=&t^{-\delta}\left(C_{\alpha>1}-\frac{(\alpha-1)\delta}{2\alpha d}  \left(\frac{x}{t^{\gamma}}\right)^2\right)_+^{\frac{1}{\alpha-1}},
\label{sec4 : eq3}  
\feqr
where
\beqr
\delta&=&d\gamma, \quad \gamma=\frac{1}{(2+d(\alpha-1))}, \label{sec4 : eq3a} \\
C_{\alpha<1}&=&\left(\frac{B(\frac{d}{2},\frac{1}{1-\alpha}-\frac{d}{2})|S_{d-1}|}{2|\kappa|^{\frac{d}{2}}}\right)^{2\gamma (1-\alpha)}, \label{sec4 : eq4} \\
C_{\alpha>1}&=&\left(\frac{2|\kappa|^{\frac{d}{2}}}{B(\frac{d}{2},\frac{\alpha}{\alpha-1})|S_{d-1}|}\right)^{2\gamma (\alpha-1)} \quad \textrm{and} \label{sec4 : eq5} \\
|\kappa|&=&\frac{|1-\alpha|}{2\alpha d}\delta
\label{sec4 : eq6}. 
\feqr 
The $d$-dimensional time-dependent functions
\beqr
f(x,t)_{\alpha<1}&=& \frac{1}{\mu_2^{d}}A_{\alpha<1} \left(1+ \beta \frac{x^2}{\mu_2^2(t)}\right)^{-\frac{1}{1-\alpha}}\quad \textrm{where} \quad \beta\equiv \beta(\alpha,d) =\frac{|1-\alpha|}{\left(2\alpha-d(1-\alpha)\right)}\label{sec4 : eq7}\\
&& \textrm{and}  \quad A_{\alpha<1}=\beta^{\frac{d}{2}}\frac{2}{|\mathbb{S}_{d-1}|B(\frac{d}{2},\frac{1}{1-\alpha}-\frac{d}{2})} \quad \textrm{if} \quad \frac{d}{d+2}<\alpha<1,
\label{sec4 : eq8}
\feqr
and
\beqr
f(x,t)_{\alpha>1}&=& \frac{1}{\mu_2^{d}}A_{\alpha>1} \left(1- \beta \frac{x^2}{\mu_2^2(t)}\right)^{\frac{1}{\alpha-1}}_+ \label{sec4 : eq9} \\
&&\textrm{where}  \quad A_{\alpha>1}=\beta^{\frac{d}{2}}\frac{2}{|\mathbb{S}_{d-1}|B(\frac{d}{2},\frac{\alpha}{\alpha-1})} \quad \textrm{if} \quad \alpha>1
\label{sec4 : eq10}
\feqr  
derived from the optimization of R\'{e}nyi's entropy, can be shown to satisfy the following initial value problem
\beqr
K_{\alpha}\partial_{t} f(x,t)&=&\Delta f^{\alpha}(x,t), \,\, (x,t)\in \mathbb{R}^d\times (0,\infty) \non \\
\lim_{t\rightarrow 0^+}\left(\int_{\mathbb{R}^d} f(x,t) dx\right)&=&1
\label{sec4 : eq11}
\feqr
provided that $\gamma \equiv \gamma(\alpha,d)=1/(2+d(\alpha-1))$ and the coefficient is given by 
\beqr
K_{\alpha \lessgtr 1}&=&\frac{2\alpha(2+d(\alpha-1))}{(2\alpha+d(\alpha-1))} A^{\alpha-1}_{\alpha \lessgtr1},
\label{sec4 : eq12} 
\feqr  
where $\mu_2(t)=t^{\gamma}$. The presence of the ratio $x/t^{\gamma}$ is implied by the self-similar property of the solution which requires the function into the parentheses to remain invariant under the rescalings: $x\rightarrow \tilde{x}=\lambda^{\gamma} x$ and $t\rightarrow \tilde{t}=\lambda t$. Therefore in general, the p.d.f. maximizing the R\'{e}nyi entropy is a solution of an appropriately constructed difussion equation problem.
\par We plot the FME and PME solutions of \rf{sec4 : eq1}  versus the solutions of \rf{sec4 : eq11}.
\begin{figure}[h]
\centering
\includegraphics[scale=1.0]{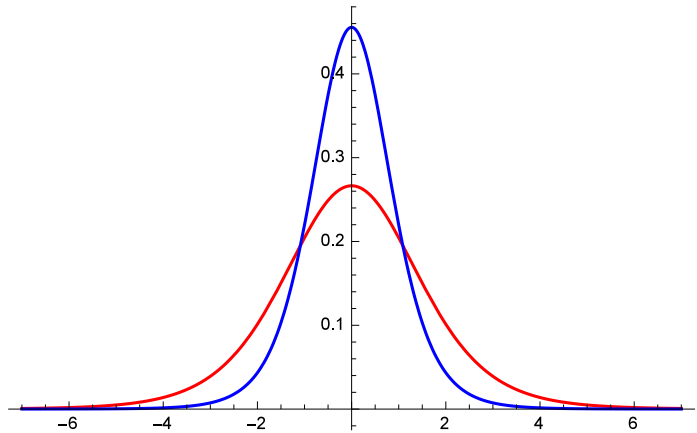} 
\caption{A snapshot at $t=1$ of the FME solution of \rf{sec4 : eq1} (red line) and \rf{sec4 : eq11} (blue line) initial value problems, with parameters: $d=1$, $\alpha=3/4$.} 
\label{sec4 : fig1}
\end{figure}
\par In the fast diffusion case $\|u\|_{\infty}<\|f\|_{\infty}$ since $C_{\alpha<1}<A_{\alpha<1}, \, \, \forall \frac{d}{d+2}<\alpha<1$ as one may prove using \rf{apA : eq14}.
\newpage
\begin{figure}[h]
\centering
\includegraphics[scale=1.0]{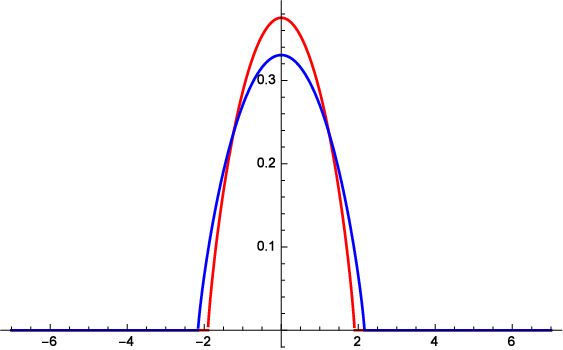} 
\caption{A snapshot at $t=1$ of the PME solution of \rf{sec4 : eq1} (red line) and \rf{sec4 : eq11} (blue line) initial value problems, with parameters: $d=1$, $\alpha=2.2$.} 
\label{sec4 : fig2}
\end{figure}
\par
In the porous medium regime $\|u\|_{\infty}<\|f\|_{\infty}$ up to the threshold value $\alpha_{\textrm{th.}}(d=1)=1.8268$ for which $C_{\alpha>1}=A_{\alpha>1}$ and then $\|u\|_{\infty}>\|f\|_{\infty}, \, \, \forall \alpha>1.8268$. The threshold value $\alpha_{\textrm{th.}}$, which depends on the dimension $d$, is determined arithmetically. 
%%%%%%%%%%%%%%%%%%%%%%%%%%%%%%%%%%%%%%%%%%%%%%%%%%%%%%%%%%%%%%%%%%%%%%%%%%%%%%%%%%%%%%%%%%%%%%%%%%
%%%%%%%%%%%%%%%%%%%%%%%%%%%%%%%%%%%%%%%%%%%%%%%%%%%%%%%%%%%%%%%%%%%%%%%%%%%%%%%%%%%%%%%%%%%%%%%%%%
\section{The concavity of R\'{e}nyi's entropy power}
\label{sec4}
\begin{definition}
Let $f : \, \Omega\times (0,\infty)\subset\mathbb{R}^d\times (0,\infty)\rightarrow \mathbb{R}^d_+\times (0,\infty)$ be a p.d.f. in $\mathcal{D}_{\alpha}$. The $\alpha$-weighted Fisher information of $f$ is defined as
\beqr
I_{\alpha}[f](t)=\frac{1}{\int_{\Omega}f^{\alpha}d\mu} \int_{\Omega} \frac{|\nabla f^{\alpha}|^2}{f}d\mu,
\label{sec5 : eq1}
\feqr
while the entropy power of $f$ associated to the R\'{e}nyi entropy $H_{\alpha}$ is defined as
\beqr
\mathcal{N}_{\alpha}[f](t)=\left\{ \begin{array}{lcc} e^{\left(\frac{2}{d}+(\alpha-1)\right)H_{\alpha}[f](t)}=\left(\int f^{\alpha}d\mu\right)^{-1} \left(\int f^{\alpha}d\mu\right)^{\frac{2}{d(1-\alpha)}}, & 0<\alpha<\infty, & \alpha\neq 1 \\ e^{\frac{2}{d}H_1[f](t)}, & \alpha=1 &\\
\mu(\{x : f(x,t)>0\})^{\frac{2}{d}-1}, & \alpha=0. &\end{array}\right.
\label{sec5 : eq2}
\feqr  
\end{definition}
\begin{proposition}
Let $\Omega=\mathbb{E}^d$ be the d-dimensional Euclidean space. The entropy power $\mathcal{N}_{\alpha}$ is a concave function of $t,\, \forall t\in(0,\infty)$ provided that :
\beqr
 \frac{d^2H_{\alpha}[f](t)}{dt^2}\leq \left\{ \begin{array}{lc} -\frac{d}{(2+d(\alpha-1))}t^{-2} & \textrm{if} \quad \frac{d}{d+2}<\alpha<1 \\ &\\
-\frac{d}{(2\alpha-1)^2(2+d(\alpha-1))}t^{-2}& \textrm{if} \quad \alpha>1, \end{array}\right.
\label{sec5 : eq3}
\feqr
where the right hand side of the inequality represents the contributions from the global maximum of $H_{\alpha}$.
\end{proposition}
\textbf{Proof}\\
Using the nonlinear diffusion equation, a straightforward calculation reveals the connection between $dH_{\alpha}/dt$ and $I_{\alpha}$ expressed by the relation
\beqr
\frac{dH_{\alpha}[f](t)}{dt} = K_{\alpha}^{-1} I_{\alpha}[f](t).
\label{sec5 : eq4}
\feqr
The entropy power is a concave function of time iff
\beqr
\frac{d^2 \mathcal{N}_{\alpha}[f](t)}{dt^2}\leq 0 \Longleftrightarrow \left(\frac{2}{d}+(\alpha-1)\right)\left(\frac{d^2 H_{\alpha}[f](t)}{dt^2}+\left(\frac{2}{d}+(\alpha-1)\right)\left(\frac{dH_{\alpha}[f](t)}{dt}\right)^2\right)\mathcal{N}_{\alpha}[f](t)\leq 0
\label{sec5 : eq5}
\feqr 
or, equivalently when
\beqr
\frac{d^2 H_{\alpha}[f](t)}{dt^2}=K_{\alpha}^{-1}\frac{dI_{\alpha}[f](t))}{dt}\leq -\left(\frac{2}{d}+(\alpha-1)\right)\left(\frac{dH_{\alpha}[f](t)}{dt}\right)^2.
\label{sec5 : eq6}
\feqr
Next, we establish the identity
\beqr
\frac{dH_{\alpha}[f](t)}{dt} = \frac{\alpha^2K_{\alpha}^{-1}}{(1-2\alpha)\int_{\mathbb{R}^d} f^{\alpha}dx} \int_{\mathbb{R}^d} f^{2\alpha -1} \Delta\left( \ln f\right) dx.
\label{sec5 : eq7}
\feqr
To do so we rewrite the integral as
\beqr
\int_{\mathbb{R}^d} \frac{|\nabla f^{\alpha}|^2}{f}dx&=& \int_{\mathbb{R}^d} f^{\alpha-1} \nabla f^{\alpha}\cdot \nabla (\ln f^{\alpha}) dx \non \\
&=&-\alpha \int_{\mathbb{R}^d} f^{\alpha} \left(\nabla f^{\alpha-1}\cdot \nabla (\ln f)+ f^{\alpha-1}\Delta (\ln f) \right)\non \\
&=& \frac{\alpha(1-\alpha)}{(2\alpha-1)}\int_{\mathbb{R}^d} \nabla f^{2\alpha-1}\cdot \nabla (\ln f)dx -\alpha \int_{\mathbb{R}^d} f^{2\alpha -1}\Delta (\ln f) dx \non \\
&=& \frac{\alpha^2}{(1-2\alpha)} \int_{\mathbb{R}^d} f^{2\alpha -1} \Delta (\ln f) dx.
\label{sec5 : eq8}
\feqr
The first time derivative of the R\'{e}nyi entropy satisfies the following upper bounds
\beqr
\frac{dH_{\alpha}[f](t)}{dt} \leq \left\{\begin{array}{lc} \frac{d}{(2+d(\alpha-1))}t^{-1} & \textrm{if} \quad \frac{d}{d+2}<\alpha<1 \\
&\\
\frac{d}{(2\alpha-1)(2+d(\alpha-1))}t^{-1}& \textrm{if} \quad \alpha>1.\end{array}\right.
\label{sec5 : eq9}
\feqr 
To prove this note that the term $\Delta (\ln \hat{f})$ is given by
\beqr
\Delta (\ln \hat{f}(x,t)) =-\frac{1}{|1-\alpha|}\frac{2\beta}{t^{2\gamma}}\times\Biggl\{\begin{array}{lc} \left(1+\beta\frac{x^2}{t^{2\gamma}}\right)^{-2}\left[d+(d-2)\beta \frac{x^2}{t^{2\gamma}}\right] & \textrm{if} \quad \frac{d}{d+2}<\alpha<1 \\
& \\
\left(1-\beta\frac{x^2}{t^{2\gamma}}\right)^{-2}\left[d-(d+2)\beta \frac{x^2}{t^{2\gamma}}\right] & \textrm{if} \quad \alpha>1\end{array}
\label{sec5 : eq10}
\feqr
and therefore its contribution to the integral is
\beqr
\int_{\mathbb{R}^d} \hat{f}^{2\alpha-1} \Delta (\ln \hat{f})dx&=& -\frac{1}{|1-\alpha|}\frac{2\beta^{1-\frac{d}{2}} |\mathbb{S}_{d-1}|}{t^{2\gamma(1+d(\alpha-1))}} \non \\
&\times&\!\!\!\!\! \left\{\begin{array}{lcc} \!\!\!\!A_{\alpha<1}^{2\alpha-1} \left[dB(\frac{d}{2},\frac{1}{1-\alpha}-\frac{d}{2})+(d-2)B(1+\frac{d}{2},\frac{\alpha}{1-\alpha}-\frac{d}{2})\right] & \textrm{if} & \frac{d}{d+2}<\alpha<1 \\
& \\
\!\!\!\!A_{\alpha>1}^{2\alpha-1} \left[dB(\frac{d}{2},\frac{\alpha}{\alpha-1})-(d+2)B(1+\frac{d}{2},\frac{\alpha}{\alpha-1}+1)\right] & \textrm{if} & \alpha>1.\end{array}\right.
\label{sec5 : eq11}
\feqr
Also,
\beqr
\frac{\alpha^2}{(1-2\alpha)} \frac{K^{-1}_{\alpha}}{\int_{\mathbb{R}^d} f^{\alpha}(x,t)dx}=\frac{\alpha}{2(1-2\alpha)}\frac{(2\alpha+d(\alpha-1))}{(2+d(\alpha-1))} \frac{\beta^{\frac{d}{2}}}{|\mathbb{S}_{d-1}|}t^{d\gamma(\alpha-1)}\times\left\{\begin{array}{lcc} \frac{A^{1-2\alpha}_{\alpha<1}}{B(\frac{d}{2},\frac{\alpha}{1-\alpha}-\frac{d}{2})} & \textrm{if} & \frac{d}{d+2}<\alpha<1 \\
&\\
\frac{A^{1-2\alpha}_{\alpha>1}}{B(\frac{d}{2},\frac{\alpha}{\alpha-1}+1)}& \textrm{if} & \alpha>1.\end{array}\right.
\label{sec5 : eq12}
\feqr
Substituting \rf{sec5 : eq9} into \rf{sec5 : eq6} we recover the expected result. \hfill\(\Box\) \\
Note that in the $\alpha \rightarrow 1^{-}$ limit we reproduce the well-known result valid for the Shannon's entropy.
\par Next we prove the concavity of R\'{e}ny's power entropy on a different setting. This problem was also studied in \cite{ST} but our approach leads to a condition not predicted before. We will need the following lemma.
\begin{lemma}
Let $G$ be defined as
\beqr
G_{\alpha}[f](t)=\int_{\mathbb{R}^d} \frac{|\nabla f^{\alpha}|^2}{f} dx=\int_{\mathbb{R}^d} f|\nabla v|^2 dx \quad \textrm{where} \quad v=\frac{\alpha}{\alpha-1}f^{\alpha-1}
\label{sec5 : eq13}
\feqr
then
\beqr
\frac{dG_{\alpha}[f](t)}{dt}&=&-2K_{\alpha}^{-1}\int_{\mathbb{R}^d} f^{\alpha}\left(\|\nabla \nabla v\|^2+(\alpha-1)(\Delta v)^2\right) dx \label{sec5 : eq14a}\\
\frac{dI_{\alpha}[f](t)}{dt}&=&(1-\alpha)K_{\alpha}^{-1}I_{\alpha}^2+\frac{1}{\int_{\mathbb{R}^d}f^{\alpha}dx}\frac{dG_{\alpha}[f](t)}{dt}.
\label{sec5 : eq14b}
\feqr
\end{lemma} 
\textbf{Proof}\\
Using the porous medium equations for $f, v$ as well as the relation $\nabla f^{\alpha}=f\nabla v$ we have
\beqr
\frac{dG[f](t)}{dt} &=& \int_{\mathbb{R}^d} \left(f_t|\nabla v|^2+2f_t\nabla v(x,t)\cdot \nabla v_t\right) dx, \,\, v_t=\frac{\partial v}{\partial t} \non \\
&=& K_{\alpha}^{-1}\int_{\mathbb{R}^d} \left(\Delta f^{\alpha}|\nabla v|^2+2 f \nabla v \cdot \nabla \left((\alpha-1)v\nabla v + |\nabla v|^2 \right)\right) dx \non \\
&=& K_{\alpha}^{-1}\int_{\mathbb{R}^d} \left(\Delta f^{\alpha}|\nabla v|^2+2(\alpha-1)\left(f|\nabla v|^2 \Delta v+fv \nabla v \cdot \nabla \Delta v \right)+2f\nabla v\cdot \nabla |\nabla v|^2 \right) dx\non \\
&=& K_{\alpha}^{-1}\int_{\mathbb{R}^d} \left(\Delta f^{\alpha}|\nabla v|^2+2(\alpha-1)\nabla f^{\alpha}\cdot \nabla v \Delta v + 2\alpha f^{\alpha}\nabla v \cdot \nabla \Delta v + 2 \nabla f^{\alpha}\cdot \nabla |\nabla v|^2\right) dx\non \\
&=& K_{\alpha}^{-1}\int_{\mathbb{R}^d} \!\!\!\!\left(\left(\Delta f^{\alpha}-2\Delta f^{\alpha}\right)|\nabla v|^2-2(\alpha-1)f^{\alpha}\left((\Delta v)^2 +\nabla v \cdot \nabla \Delta v \right)+2\alpha f^{\alpha}\nabla v \cdot \nabla \Delta v \right)dx \non \\
&=& K_{\alpha}^{-1}\int_{\mathbb{R}^d} \left(f^{\alpha}\left(2\nabla v \cdot \nabla \Delta v-\Delta(|\nabla v|^2)\right)-2 (\alpha-1)f^{\alpha}(\Delta v)^2\right)dx\non \\
&=& -2K_{\alpha}^{-1}\int_{\mathbb{R}^d} f^{\alpha}\left(\|\nabla \nabla v\|^2+(\alpha-1)(\Delta v)^2\right)dx,
\label{sec5 : eq15}
\feqr 
where partial integrations in the fourth and fifth equalities have been used. Also in the last step the Bochner's formula in Euclidean space has been applied. Relation \rf{sec5 : eq14b} is proved using \rf{sec5 : eq14a}. \hfill\(\Box\)
\begin{theorem}
The R\'{e}nyi entropy power, for self-similar solutions, is concave in t provided that the following inequality is satisfied
\beqr
\int_{\mathbb{R}^d}\left(f^{\alpha}(\xi)-C f^{2\alpha-1}(\xi)\right)d\xi\geq 0 ,\quad \xi=\frac{x}{t^{\gamma}}, \quad C=\frac{1}{\alpha^2 (1+d(\alpha-1))}, 
\label{sec5 : eq16}
\feqr
where the equality holds for $\alpha=1$.
\end{theorem} 
\textbf{Proof}\\
The R\'{e}nyi entropy power is concave in t iff the condition\rf{sec5 : eq6} holds which can be written equivalently as
\beqr
I^2_{\alpha}[f](t) &\leq& d \left(\int_{\mathbb{R}^d}f^{\alpha}dx\right)^{-1} \int_{\mathbb{R}^d}f^{\alpha}\left(\|\nabla \nabla v\|^2+(\alpha-1)(\Delta v)^2\right)dx \non \\
&\leq& d(1+d(\alpha-1))\left(\int_{\mathbb{R}^d}f^{\alpha}dx\right)^{-1} \int_{\mathbb{R}^d}f^{\alpha}\|\nabla \nabla v\|^2 dx\non \\
&\leq& d(1+d(\alpha-1)) J_{\alpha}[f](t),
\label{sec5 : eq17}
\feqr
where $\|\nabla \nabla v\|^2=\| \textrm{Hess}\, v\|^2=\sum_{i,j=1}^d (\partial_{ij}v)^2$ and we used the identity $\| \textrm{Hess}\, v\|^2=\textrm{Tr} \left((\textrm{Hess}\, v)^2\right)\geq \left(\textrm{Tr} (\textrm{Hess}\, v)\right)^2/d$. Expanding $\|\cdot\|^2$ and writing it in terms of $\ln f$ the last integral can be casted into the form 
\beqr
\int_{\mathbb{R}^d}f^{\alpha}\|\nabla \nabla v\|^2 dx=\alpha^2 \int_{\mathbb{R}^d}f^{2\alpha-1} \sum_{i,j=1}^d \left((\alpha-1)(\partial_i \ln f )(\partial_j \ln f) + \partial_{ij}^2 \ln f \right)^2 dx,
\label{sec5 : eq18}
\feqr
which by applying the Cauchy-Schwarz inequality we get that
\beqr
\int_{\mathbb{R}^d}f^{\alpha}\|\nabla \nabla v\|^2 dx\geq  \frac{\alpha^2}{d} \left(\int_{\mathbb{R}^d}f^{2\alpha-1}dx\right)^{-1}\left(\int_{\mathbb{R}^d}f^{2\alpha-1}\left((\alpha-1)|\nabla \ln f|^2 + \Delta \ln f \right)dx\right)^2.
\label{sec5 : eq19}
\feqr
Using the identity
\beqr
\frac{\Delta f^{\alpha}}{f^{\alpha}}=|\nabla \ln f^{\alpha}|^2+\Delta \ln f^{\alpha}
\label{sec5 : eq20}
\feqr
we eliminate the $|\nabla \ln f|^2$ term from the previous relation and get
\beqr
J_{\alpha}[f](t)\geq \frac{\alpha^2}{d} \frac{\left(\int_{\mathbb{R}^d}f^{\alpha}\right)}{\left(\int_{\mathbb{R}^d}f^{2\alpha-1}\right)} I^2_{\alpha}(f)\geq \frac{1}{d(1+d(\alpha-1))} I^2_{\alpha}(f).
\label{sec5 : eq21}
\feqr
\hfill\(\Box\)\\
In 1979, Aronson and B\'{e}nilan obtained a second-order differential inequality of the form \cite{AB}
\beqr
\sum_{i=1}^d \frac{\partial}{\partial x_i}\left(\alpha f^{\alpha-2}\frac{\partial f}{\partial x_i} \right)=\Delta\left(\frac{\alpha}{\alpha-1}f^{\alpha-1}\right)\geq -\frac{d}{\left(d(\alpha-1)+2\right)t}, \quad \alpha>\alpha_c:=1-\frac{2}{d}, 
\label{sec5 : eq23}
\feqr
which applies to all positive smooth solutions of the porous medium equation defined on the whole Euclidean space\footnote{In $d=1$ the restriction is $\alpha>0$ for general solutions.}. 
In 1986 Li-Yau studied a heat type flow \cite{LY} on complete Riemannian manifolds $(M,g)$ with nonnegative Ricci scalar  and for any positive function $f$ on M and any $t>0$ they arrived at the following sharp lower bound 
\beqr
\Delta(\ln f)\geq -\frac{d}{2t}, 
\label{sec5 : eq22}
\feqr
pointwise, where $\Delta$ is the Laplace-Beltrami operator on $M$. An extension of the Aronson and B\'{e}nilan estimate to the PME flow for all $\alpha>1$ and the FME flow for $\alpha\in (\alpha_{c},1)$ on complete Riemannian manifolds with Ricci scalar bounded from below was given in \cite{LNVV}. 
\par In our case these bounds change because our p.d.f. differs from the solution of the PME derived from the gradient flow of the functional \cite{FO}
\beqr
E[f](t)=\frac{1}{1-\alpha} \int_{\mathbb{R}^d} f^{\alpha}(x,t) dx.
\label{sec5 : eq23}
\feqr
%%%%%%%%%%%%%%%%%%%%%%%%%%%%%%%%%%%%%%%%%%%%%%%%%%%%%%%%%%%%%%%%%%%%%%%%%%%%%%%%%%%%%
%%%%%%%%%%%%%%%%%%%%%%%%%%%%%%%%%%%%%%%%%%%%%%%%%%%%%%%%%%%%%%%%%%%%%%%%%%%%%%%%%%%%%
\section{Conclusions}
\label{sec5}
\par In this article we have proved that the p.d.f.'s that maximize R\'{e}nyi's entropy under the conditions of finite variance and of zero or non-zero mean are given by a one-parameter family of functions which belong to $S(\mathbb{R}^d)$ for $\alpha<1$ and to $C^l_0(\mathbb{R}^d)$ for $\alpha>1$. This one-parameter family of functions is a global maximum of the entropy and satisfies the non-linear diffusion equation \rf{sec4 : eq11}. The $\mathcal{L}_{\infty}$ norms of these solutions when compared to the corresponding ones derived from the fast and porous medium diffusion equation initial value problem \rf{sec4 : eq1}, they appear to behave in a particular way as seen in Figures \rf{sec4 : fig1} and \rf{sec4 : fig2}.
\par If one considers finite even moments of the random variable $X$, as constraints, and try to solve the corresponding maximization problem then the $S(\mathbb{R})$ solution exists whenever $\hat{f}^{\alpha-1}$ is a complete polynomial of even degree, or equivalently, when all the coefficients $\tilde{\lambda}_{2k+1}$ vanish. In those cases there exists the possibility not to have interception points with the x-axis since the roots come into conjugate complex pairs. The compactly supported solution, under certain conditions, can always be determined.   
\par The concavity of entropy power holds whenever the second time derivative of the entropy varies according to \rf{sec5 : eq3} or the function $f^{\alpha}-Cf^{2\alpha-1}$ belongs to $L_1(\mathbb{R}^d)$. It would be more appealing to have a deeper understanding of the origin of the later constraint which at this stage seems to be a requirement for consistency.
%%%%%%%%%%%%%%%%%%%%%%%%%%%%%%%%%%%%%%%%%%%%%%%%%%%%%%%%%%%%%%%%%%%%%%%%%%%%%%%%%%%
%%%%%%%%%%%%%%%%%%%%%%%%%%%%%%%%%%%%%%%%%%%%%%%%%%%%%%%%%%%%%%%%%%%%%%%%%%%%%%%%%%%
\section{Acknowledgments}
The author would like to thank both J. K. Pachos for the fruitful discussions he had related to this project, and the Department of Physics and Astronomy of the University of Leeds for its hospitality during this visit. 
%%%%%%%%%%%%%%%%%%%%%%%%%%%%%%%%%%%%%%%%%%%%%%%%%%%%%%%%%%%%%%%%%%%%%%%%%%%%%%%%%%%
%%%%%%%%%%%%%%%%%%%%%%%%%%%%%%%%%%%%%%%%%%%%%%%%%%%%%%%%%%%%%%%%%%%%%%%%%%%%%%%%%%%
\addcontentsline{toc}{subsection}{Appendix}
\section*{Appendix}
\label{apA}
\renewcommand{\theequation}{A.\arabic{equation}}
\setcounter{equation}{0}
Using the two integral formulas \cite{GR} 
\beqr
\int_0^{\infty} x^{\mu-1}(1+x^2)^{\nu-1} dx&=&\frac{1}{2} B(\frac{\mu}{2},1-\nu-\frac{\mu}{2}), \quad \textrm{Re}\mu>0,\, \textrm{Re}(\nu+\frac{\mu}{2})<1\non \\
\int_0^1 x^{\mu-1}(1-x^{\lambda})^{\nu-1} dx&=&\frac{1}{\lambda}B(\frac{\mu}{\lambda},\nu), \quad \textrm{Re}\mu,\,\textrm{Re}\nu, \,\lambda>0, 
\label{apA : eq1}
\feqr
one can prove the following ones, used in section 3,
\begin{enumerate}
\item 
\beqr
\int_0^{\infty} x^{\mu-1}(1+x^2)^{\nu-1} dx=\frac{1}{2}\int_0^1 u^{\frac{\mu}{2}-1}(1-u)^{-\nu-\frac{\mu}{2}}=\frac{1}{2} B(\frac{\mu}{2},1-\nu-\frac{\mu}{2}),
\label{apA : eq2}
\feqr
where we made the change of variable $u=x^2/(1+x^2)$. 
\item 
\beqr
 \int_{\mathbb{S}_{d-1}} \int_{0}^{\infty} \!\!\!\!\frac{r^{d-1}}{(C(\alpha)+g(\alpha) r^2)^{\lambda(\alpha)}}  dr\, dS= |\mathbb{S}_{d-1}| \frac{C^{\frac{d}{2}-\lambda}}{2 g^{\frac{d}{2}}}B(\frac{d}{2},\lambda-\frac{d}{2}) \,\,\, \textrm{where} \,\,\, C,g,\lambda>0, \, \, d<2\lambda, 
\label{apA : eq3}
\feqr
$|\mathbb{S}_{d-1}|=\frac{2\pi^{\frac{d}{2}}}{\Gamma\left(\frac{d}{2}\right)}$ is the volume of the unit ball and $B(x,y)=\Gamma(x) \Gamma(y)/\Gamma(x+y)$ is the Euler's beta function. The constraint $d<2\lambda$ for $\lambda=1/(1-\alpha)$ allows only values
\beqr
\alpha_{c_0}=1-\frac{2}{d}<\alpha<1 \quad \textrm{for} \quad d\geq 3,
\label{apA : eq4}
\feqr
while $\alpha_{c_0}=0$ for $d=1,2$.
\item 
\beqr
 \int_{\mathbb{S}_{d-1}} \int_{0}^{\infty} \!\!\!\!\!\! \frac{r^{d+1}}{(C(\alpha)+g(\alpha) r^2)^{\lambda(\alpha)}}  dr\, dS= |\mathbb{S}_{d-1}| \frac{C^{\frac{d}{2}+1-\lambda}}{2 g^{1+\frac{d}{2}}}B(\frac{d}{2}+1,\lambda-\frac{d}{2}-1) \,\, \textrm{where} \,\, d+2<2\lambda. 
\label{apA : eq5}
\feqr
The constraint between $d, \, \lambda$ becomes
\beqr
\alpha_{c_2}=1-\frac{2}{d+2}<\alpha<1 \quad \textrm{for} \quad d\geq 1.
\label{apA : eq6}
\feqr
\item The last formula is
\beqr
&& \int_{\mathbb{S}_{d-1}} \!\!\int_{0}^{\infty} \!\!\!\!\! r^{d-1}(C(\alpha)-g(\alpha) r^2)_{+}^{k(\alpha)}  dr\, dS= |\mathbb{S}_{d-1}| \frac{1}{2} g(\alpha)^{-\frac{d}{2}} C(\alpha)^{\frac{d}{2}+k(\alpha)}B(\frac{d}{2},k(\alpha)+1) \non \\
&& \textrm{when} \quad k(\alpha)>0.
\label{apA : eq7}
\feqr 
\end{enumerate}
Formulas \rf{apA : eq3}-\rf{apA : eq4} are direct consequences of \rf{apA : eq1} and \rf{apA : eq2}.
Another integral formula \cite{GR} used in section 4 is
\beqr
\int_{\mathbb{R}}\frac{x^m}{(ax^2+2\beta x +c)^n}&=&\frac{(-1)^m\pi a^{n-m}b^m}{(2n-2)!!(ac-b^2)^{n-\frac{1}{2}}} \non \\
&\times& \sum_{k=0}^{[m/2]}\left(\begin{array}{c}m\\ 2k \end{array}\right)(2k-1)!! (2n-2k-3)!! \left(\frac{ac-b^2}{b^2}\right)^k \non \\
&& \textrm{where} \quad ac-b^2>0, \,\, 0\leq m\leq 2(n-1) 
\label{apA : eq8}
\feqr
and $m, n$ are nonnegative, positive integers respectively. 
\par Identities involving the Euler's beta function
\begin{enumerate}
\item
\beqr
\lim_{\rho \rightarrow \infty} \left(\rho^{\frac{s}{2}} B(\frac{s}{2}, \rho-\frac{s}{2})\right)=\Gamma\left(\frac{s}{2}\right)
\label{apA : eq9}
\feqr
\textbf{Proof}\\
The large asymptotic expansion of the ratio of gamma functions \cite{TE}, is given by
\beqr
\frac{\Gamma(\rho+a)}{\Gamma(\rho+b)}=\rho^{a-b}\left(1+\frac{(a-b)(a+b-1)}{\rho}+O(\rho^{-2})\right).
\label{apA : eq10}
\feqr
Substituting  the values $a=-s/2$, $b=0$ in the previous expression and taking the limit $\rho \rightarrow \infty$ we recover the desired result.\\
\item
\beqr
d\frac{B(\frac{d}{2},\frac{1}{1-\alpha}-\frac{d}{2})}{B(\frac{d}{2},\frac{\alpha}{1-\alpha}-\frac{d}{2})}+(d-2)\frac{B(1+\frac{d}{2},\frac{\alpha}{1-\alpha}-\frac{d}{2})}{B(\frac{d}{2},\frac{\alpha}{1-\alpha}-\frac{d}{2})}&=&\frac{d}{\alpha}(2\alpha-1)\non \\
d\frac{B(\frac{d}{2},\frac{\alpha}{\alpha-1})}{B(\frac{d}{2},\frac{\alpha}{\alpha-1}+1)}-(d+2)\frac{B(1+\frac{d}{2},\frac{1}{\alpha-1}+1)}{B(\frac{d}{2},\frac{\alpha}{\alpha-1}+1)}&=&\frac{d}{\alpha}.
\label{apA : eq11}
\feqr
These are direct consequences of the definition of the Euler's beta function and the gamma's function property $\Gamma(x+1)=x\Gamma(x)$.
\item 
\beqr
B(\frac{1}{2},n-\frac{1}{2})&=& \frac{(2n-3)!!}{(2n-2)!!}\pi \non \\
B(k+\frac{1}{2},n-k-\frac{1}{2})&=& (2k-1)!! \frac{(2n-2k-3)!!}{(2n-2)!!}\pi, \quad k\in \mathbb{Z}^+ \non \\
B(\frac{1}{2},n+1)&=& 2^{2n+1}B(n+1,n+1)
\label{apA : eq12}
\feqr
\textbf{Proof}\\
Use the doubling formula for gamma functions and $\Gamma(n)$ in terms of the double factorial
\beqr
\Gamma\left(n+\frac{1}{2}\right)&=& \frac{\sqrt{\pi}}{2^n}(2n-1)!! \non \\
\Gamma(n)&=&\frac{(2n-2)!!}{2^{n-1}}
\label{apA : eq13}
\feqr 
\end{enumerate}
\begin{proposition}
The function $\Gamma(x)/\Gamma(x+c)$ with $x,c \, \in\mathbb{R}^+$ is decreasing in $x$.
\label{apA : eq14}
\end{proposition}
\textbf{Proof}
It is enough to prove that 
\beqr
\frac{d}{dx} \ln\biggl(\frac{\Gamma(x)}{\Gamma(x+c)} \biggr)=\psi(x)-\psi(x+c)<0.
\label{apA : eq15}
\feqr
A straightforward computation gives
\beqr
-\ln\biggl(1+\frac{c}{x} \biggr) -\frac{c}{x(x+c)}\leq \psi(x)-\psi(x+c) \leq -\ln\biggl(1+\frac{c}{x} \biggr) -\frac{c}{2x(x+c)}<0
\label{apA : eq16}
\feqr
by using the inequality
\beqr
\ln x -\frac{1}{x}\leq \psi(x) \leq \ln x -\frac{1}{2x}.  
\label{apA : eq17}
\feqr
%%%%%%%%%%%%%%%%%%%%%%%%%%%%%%%%%%%%%%%%%%%%%%%%%%%%%%%%%%%%%%%%%%%%%%%%%%%%%%%%%%%%%%%%%%%%%%%%
%\textbf{Conflicts of Interest}\\

%The author declares no conflict of interest.

%%%%%%%%%%%%%%%%%%%%%%%%%%%%%%%%%%%%%%%%%%%%%%%%%%%%%%%%%%%%%%%%%%%%%%%%%%%%%%%%%%%%%%%%%%
%%%%%%%%%%%%%%%%%%%%%%%%%%%%%%%%%%%%%%%%%%%%%%%%%%%%%%%%%%%%%%%%%%%%%%%%%%%%%%%%%%%%%%%%%%
\bibliographystyle{plain}

\end{document}